\documentclass[11pt,a4paper]{article}
\usepackage[dvips]{graphicx}
\usepackage[normal,small,bf]{caption2} %%%%%     Legendes des figures

\usepackage{vmargin}
\setmarginsrb{3.0cm}{3.7cm}{3.0cm}{3.0cm}{0cm}{-0.8cm}{1.0cm}{1.5cm}
\usepackage{fancyheadings}
\newcommand{\be}{\begin{equation}}
\def\bfig#1{\begin{figure}[#1] \setcaptionwidth{12cm} \renewcommand{\captionlabeldelim}{.\ } \centering  \addtolength{\fboxsep}{5mm} }
\newcommand{\diff}{{\rm d}}
\newcommand{\dm}{\langle x^2\rangle}
\newcommand{\dt}{\langle \tau^2\rangle}
\newcommand{\ee}{\end{equation}}
\def\efig{\end{figure} \pagebreak[1]}
\newcommand{\ie}{i.e. }
\newcommand{\kB}{k_{\rm B}}
\def\kB{k_{\rm B}}

\newcommand{\pmo}{\langle x \rangle}
\newcommand{\pt}{\langle \tau \rangle}
\newcommand{\ptrap}{p_{\rm trap}}
\newcommand{\SR}{S_{\rm R}}
\newcommand{\tauexp}{\tau_{\rm exp}(N)}
\newcommand{\tl}{t_{l}}
\def\TR{T_{\rm R}}

\title{Cooling gases with L\'evy flights:
\\using the generalized central limit theorem \\ in physics 
}
\author{Fran\c{c}ois {\sc Bardou}\thanks{Email: bardou@morgane.u-strasbg.fr.
Presently visitor at : Department of Physics, University of Newcastle, {\sc
Newcastle-upon-Tyne NE1 7RU}, United Kingdom.}}
\date{}
\begin{document}
\maketitle

\begin{center}
{\it Institut de Physique et de Chimie des Mat\'eriaux de Strasbourg, \\
23 rue du Loess, F-67037 STRASBOURG Cedex, France}
\end{center}

\begin{center}
\begin{minipage}{12cm}
\textsf{\small
In the last ten years, the generalized central limit theorem
established by Paul L\'evy in the thirties has been found more and more
relevant in physics. Physicists call 'L\'evy flights'
random walks for which the probability density of the jump lenghts $x$ decays
as $1/x^{1+\alpha}$ with $\alpha<2$ for large $x$.
We give here a glimpse of L\'evy flights in physics
through two examples, without going into technical details. We first 
introduce a simple toy model, the Arrhenius cascade.
We then present an important physical process, subrecoil laser cooling
of atomic gases, in which L\'evy flights play an essential role.
}
\end{minipage}
\end{center}

\chead{\small \it Mini-proceedings: Conference on 'L\'evy processes: theory and
applications'\\ 
Aarhus 18-22 january 1999, MaPhySto Publication (Miscellanea no. 11, ISSN 1398-5957),\\ O. Barndorff-Nielsen, S.E.~Graversen and T.~Mikosch (eds.).}
\thispagestyle{fancy}

\tableofcontents

\section{Introduction}

The 'usual' central limit theorem (CLT) is an essential tool in physics
and in other sciences. Indeed, one often knows the probability density 
$f(x)$ of a quantity $x$ associated with single events. Then, one wants to
derive the probability density $f(X_N)$ for a sum $X_N$
\be
X_N = \sum_{i=1}^N x_i
\ee
of a large number $N$ of such events, considered as independent\footnote{For
simplicity, any sum of independent events is called a 'L\'evy sum' below.}.
In simple terms, the usual CLT tells that the density of 
$(X_N-N\langle x \rangle )/\sqrt{N}$ tends to a gaussian distribution at
large $N$ and that this distribution is determined only by the average
value $\langle x\rangle$ and the second moment $\langle x^2\rangle$,
provided that these quantities are finite\footnote{If $\langle x^2
\rangle$ is finite, we say that $f(x)$ is a 'narrow' probability density.}.
The condition for applying the usual CLT (finiteness of
$\dm$) is so frequently satisfied that most physicists
implicitely believe that this theorem applies universally.

However, physical phenomena can exhibit statistical properties that are 
beyond the usual CLT. In particular, densities $f(x)$ with power law
tails:
\be
f(x) \sim \frac{1}{x^{1+\alpha}} \quad \mathrm{for} \quad x\to \infty
\label{puiss}
\ee
(with $\alpha>0$ to ensure normalizability) are simple laws that tend
to appear frequently. If $\alpha>2$, $\langle x^2\rangle$ is finite 
and the usual CLT applies. On the contrary, if $\alpha\leq2$, $\dm$ 
diverges\footnote{We then say that $f(x)$ is a 'broad' probability density.} 
and the usual CLT does not apply. 
If $\alpha\leq 1$, even $\pmo$ diverges. As stated by the 'generalized' CLT,
if $0<\alpha< 2$, the density of $X_N$ (properly renormalized) still
tends to a stable law, which is not a gaussian but a L\'evy law.
After Mandelbrot, we call 'L\'evy flight' a random walk in which the
probability density of jump lengths is given by Eq.~(\ref{puiss}) 
with $\alpha<2$.

\medskip
The generalized CLT was already available in the thirties but, surprisingly,
it has had a limited influence on physics for a long time. Most physicists
were certainly not aware of it and those who were aware probably doubted that 
infinite average values $\pmo$ or second moments $\dm$ could make sense in a
real phenomenon. Note that few cases which come under the generalized CLT
were known, but they remained isolated cases
(such as the density of first return times $\tau$ in one dimension
which decays as $1/\tau^{3/2}$ for large $\tau$'s).

In the recent years, it has been more and more recognized that the 
generalized CLT could shed an interesting light on many physical processes:
random walks in solutions of micelles~\cite{OBL90}, turbulent and 
chaotic transport~\cite{SZK93,SWS93}, 
glassy dynamics~\cite{Shl88,BoD95}, diffusion of spectral lines in disordered
solids~\cite{ZuK94}, thermodynamics~\cite{TLS95,Tsa95,ZaA95}, 
granular flows~\cite{BoC97}, laser trapped ions~\cite{MEZ96,KSW97} ... 
For reviews, see \cite{Shl88,BoG90,KSZ96,Tsa97}. 

\smallskip
The interest of the physics community
for the generalized CLT seems to be stimulated by two important arguments.

First, the phenomena obeying only the generalized form of the CLT, \ie
those with asymptotic power laws for $f(x)$ with $\alpha < 2$, exhibit
a statistical behaviour which is markedly different from the behaviour
of the phenomena obeying the usual CLT. It is thus important to identify
whether a physical process comes under the generalized form of the CLT
(in the first place to avoid the use of natural but irrelevant concepts,
such as the average value, derived from the usual CLT). This is illustrated
in section~2 with the simple model of the Arrhenius cascade.

Second, the generalized CLT provides an efficient tool for the quantitative 
study of some physical
problems. This is illustrated in section~3 with a L\'evy flight theory of
laser cooling of atomic gases. In this case, it is worth noting that the 
ideas derived from the generalized CLT have had practical consequences,
leading to more efficient cooling strategies and to record low temperatures.

Finally, the generalized CLT also provides a useful qualitative insight
for some random walks even when it is not strictly valid. This is discussed
briefly in section~4.

\medskip
{\it \small Note that this contribution presents the point of view of 
a physicist and, as such, might not be rigorous on all mathematical aspects.}

\section{The Arrhenius cascade}

We present the model of the Arrhenius cascade in section~2.1, which is shown
to exhibit an unexpected statistical behaviour in section~2.2, analyzed
with the generalized CLT in section~2.3. This toy model presents generic
effects of the generalized CLT in physics.

\subsection{The model}

We consider a physical system placed in the potential landscape schematized in
figure~1 and submitted to thermal fluctuations. This model is called here the
Arrhenius cascade. It is inspired from studies of disordered systems, like
glasses relaxing towards low energy states~\cite{Shl88,BoD95}, 
which obey similar equations.

\bfig{!h}
\vspace*{0.5cm}
\includegraphics[height=5cm,width=8cm,angle=0]{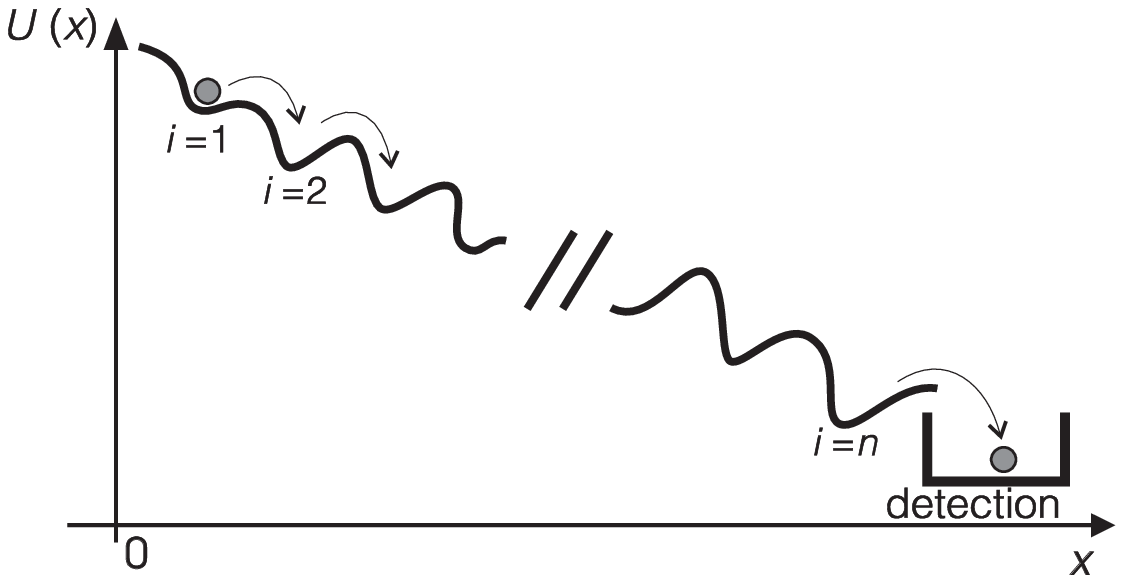}
\caption{The Arrhenius cascade.}
\label{fig1}
\end{figure}

The potential $U(x)$ is a tilted random 'washboard'. It presents $n$ local
minima or 'wells', labelled by $i$, separated from the next minimum $i+1$ by
a potential 'hill' of random height $U_i$ (the $U_i$'s are independent
variables). We assume an exponential probability density $f(U)$ for the hills
$U_i$:
\be
f(U)=\frac{1}{U_0} \; e^{-U/U_0} \, ,
\label{exp}
\ee
where $U_0$ is the average height of the potential hills. The variable $x$ 
may be a position coordinate or any coordinate of the system.

At any time, the physical system is trapped in one of the local potential
wells. Due to thermal fluctuations, the system performs sudden jumps from one
well to the other one downwards. 
The global tilt of the potential hill $U(x)$ is
large enough to prevent the system from performing upward jumps. Therefore, the
system can only cascade downwards. The trapping time $\tau_i$ in the well $i$
is related to $U_i$ by the Arrhenius law\footnote{In a realistic model,
$\tau_i$ is not deterministically fixed by $U_i$ and the
expression~(\ref{Arrhenius}) gives only the average value of $\tau_i$. Taking into
account the fluctuations of $\tau_i$ for a given $U_i$ would not change
qualitatively the discussion presented here.}:
\be
\tau_i = \tau_0 \; e^{U_i/(\kB T)} \, ,
\label{Arrhenius}
\ee
where $\tau_0$ is a characteristic time, $\kB$ is the Boltzmann constant
and $T$ is the temperature. 

We consider a gedanken experiment in which the experimentalist wants to know
the average trapping time in a single well, but can only measure the time 
$t_n$ needed to go through all the $n$ wells of the system:
\be
t_n = \sum_{i=1}^{n} \tau_i \ \cdot
\ee
To reduce the measurement uncertainty, he can repeat $m$ times his measurement
of $t_n$. We assume that, between each measurement, the realization of the
disorder changes, \ie the numbers $U_1,... U_n$ change (and thus $\tau_1,...
\tau_n$ change accordingly). His estimation $\tauexp$ of the average trapping
time will therefore be:
\be
\tauexp = \frac{1}{N} \sum_{j=1}^{m} t_n = \frac{1}{N} \sum_{i=1}^{N} \tau_i \ ,
\ee
where $N=m\times n$ is the total number of explored wells and the $\tau_i$'s
are independent random variables defined by Eq.~(\ref{exp}) and
(\ref{Arrhenius}).

\subsection{Behaviour of the Arrhenius cascade}

\bfig{!h}
\includegraphics[height=8cm,width=5.5cm,angle=-90]{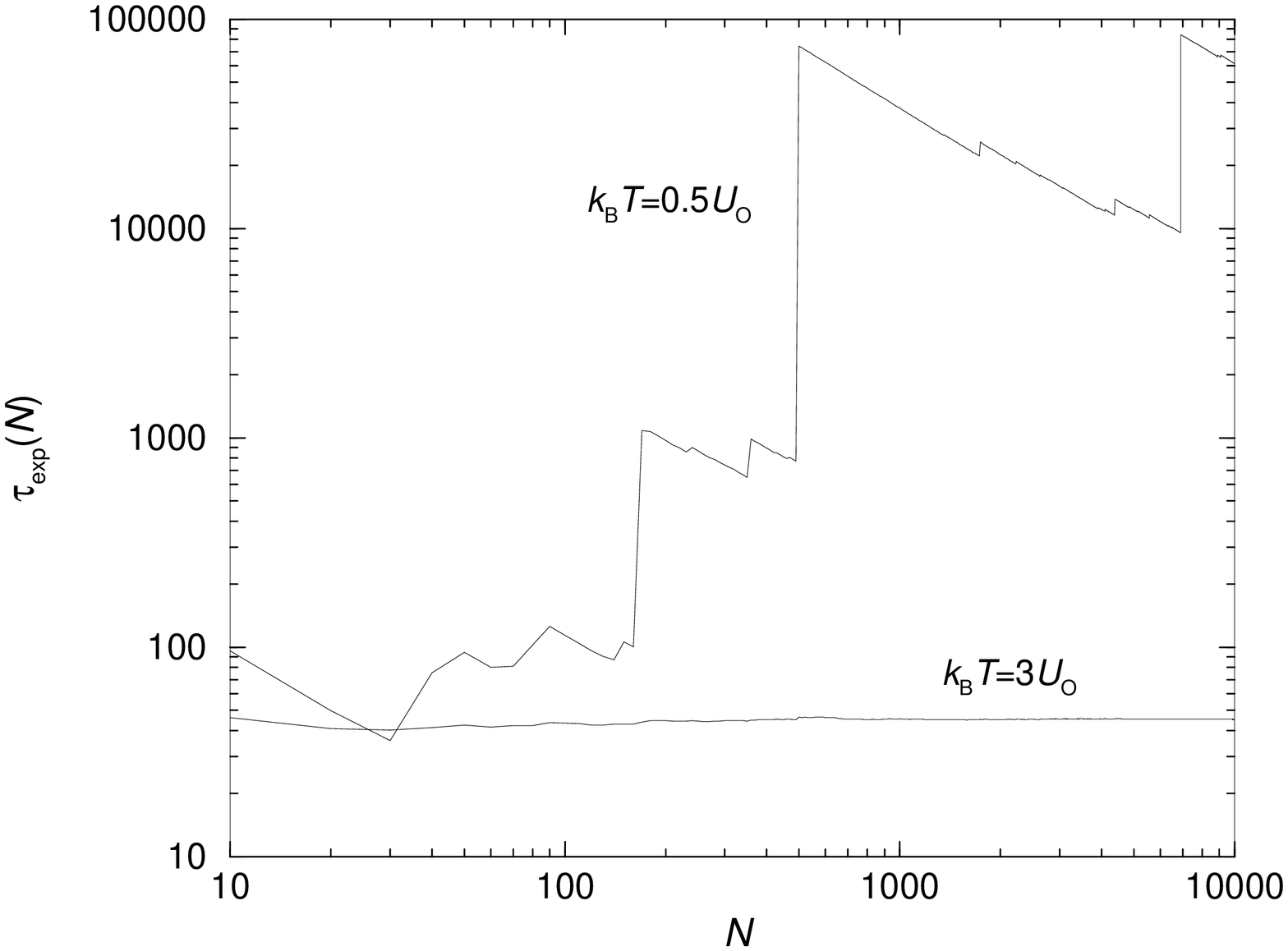}
\caption{Simulated measurements of the estimated $\tauexp$. We have chosen
$n=10$ and $\tau_0=1$. The quantity $\tauexp$ for the case 
$\kB T = 3U_0$ has been multiplied by 30 to be more visible.}
\label{fig2}
\end{figure}

Simulated measurements of $\tauexp$ are represented in figure~2 for two
different temperatures. For a temperature $T=3U_0/\kB$, $\tauexp$ converges
nicely to the average value $\pmo$ when $N$ increases, after exhibiting
reasonable fluctuations at small $N$. This is the expected, standard
behaviour. 

For a lower temperature $T=0.5U_0/\kB$, the behaviour of $\tauexp$
is markedly different. The measured $\tauexp$'s do not seem to converge 
towards any constant value but rather to diverge in a very fluctuating way
with increasing $N$. A detailed analysis would reveal that the $t_n$'s
also fluctuate very much from one measurement to the other. This unusual
statistical behaviour would puzzle most experimentalists: large fluctuations
and irreproducibility are usually considered as the indication of a problem in
the experimental procedure, arising from poorly controlled
parameters. But here, one would find that 
the experimental setup works apparently well
\footnote{A closely related situation has recently appeared in
a quantum tunneling problem. See section~4.}.

\subsection{Application of the generalized CLT}

Let us see how the generalized CLT can shed light on the previous 
observations. We first
easily calculate the probability density $f(\tau)$ of the trapping times.
It is given by the relation $f(\tau)\diff \tau = f(U) \diff U$, which leads
immediately to
\be
f(\tau) = \alpha\; \frac{\tau_0^\alpha}{\tau^{1+\alpha}} \quad
{\rm with} \quad \alpha= \frac{\kB T}{U_0} \; \cdot
\ee
Having a power law for $f(\tau)$, the probability densities of $\tauexp$ at 
large $N$ are provided for any $\alpha$ by the generalized CLT.

In the high temperature case ($T=3U_0/\kB$), we have $\alpha=3$ which is
larger than 2. Thus $\dt$ is finite and the usual CLT applies: 
$\tauexp$ is gaussianly distributed at large $N$ and tends to $\pt$.
This agrees with the observations (figure~2) and does not need further
discussion.

In the low temperature case ($T=0.5U_0/\kB$), on the contrary, 
we have $\alpha=0.5$ which is
smaller than 1. The usual CLT does not hold anymore and the specific features
of the generalized CLT will play a crucial role. The generalized CLT tells us
that we should consider the quantity $Z_N=\sum_{i=1}^{N}\tau_i/N^{1/\alpha}$ and
that the density $f(Z_N)$ tends to a L\'evy law $L_\alpha(Z_N)$ at large
$N$. This theorem has several important physical consequences: 
scaling of the L\'evy
sums, domination of the L\'evy sums by a single term and large fluctuations of
the L\'evy sums (most of these consequences are presented in~\cite{BoG90}). We only treat here the case $\alpha<1$ (and not $1\leq\alpha\leq2$)
for which the consequences of the generalized CLT depart most strongly 
from the ones of the usual CLT. These consequences are:

\begin{itemize}

\item The most probable value of the sum $T_N=\sum_{i=1}^{N}\tau_i$ 
{\it scales} as
\be
T_N \sim N^{1/\alpha}\, ,
\ee
and not as $N$ which is more usual. Practically, this implies
that the time $t_n$ spent in an Arrhenius cascade of $n$ wells does not scale
with the size $n$ of the cascade, but more rapidly due to the generalized CLT\footnote{Such 'anomalous' scaling with the system size appears in some complex 
phenomena like phase transitions near a critical point. What is striking 
here is to obtain such scaling in a very simple problem.}. 
Similarly, the experimental value
$\tauexp=\frac{1}{m\times n} \sum_{j=1}^{m} t_n$ does not tend to a constant
for a large number of measurements $m$ but diverges with $m$, as
$m^{-1+1/\alpha}$. This explains the observed behaviour in figure~2. Thus, the
size of the system and the number of measurements play a non trivial role in 
the measured values, an unusual situation in physics.

\item The notion of average value is irrelevant here since $\pt = \infty$. One
can somehow replace it by the notion of {\it typical} value, \ie most probable
value. When $\alpha<1$, the typical terms $x_i$ of a L\'evy sum
$X_N=\sum_{i=1}^{N}x_i$ are not all of the same order (as they are when $f(x)$
is a narrow density) but present a hierarchical structure. In particular,
{\it the typical largest term $x_{\rm max}$ of the sum can be shown to be of
the same order of magnitude as the sum itself
\be
X_N = \sum_{i=1}^{N}x_i \simeq x_{\rm max} \, ,
\label{domine}
\ee
however large $N$ might be (Eq.~(\ref{domine}) is valid within prefactors 
that do not depend on $N$).}
This domination of a sum by a single term, or by a small number of terms, is a
signature of L\'evy statistics in a physical problem (see figure~3 below).

It can be used cleverly in physical experiments as a 'L\'evy microscope': by
measuring a macroscopic quantity $X_N=\sum_{i=1}^{N}x_i$, one may obtain an easy
access to a microscopic information, $x_{\rm max}$, while the direct
measurement of $x_{\rm max}$ (\ie in the Arrhenius cascade, the direct study
of a single well) can be physically impossible. Such advantageous use of the
statistical domination of a single term has been made implicitly, for instance
in studies of quantum tunneling~\cite{RoB84}.

\item 
{\it A L\'evy sum $X_N=\sum_{i=1}^{N}x_i$ fluctuates as much as a single term, 
when $\alpha<1$.} This is
a direct consequence of the domination of the sum by a single term. It can
also be seen as a consequence of the fact that the tail of the L\'evy  law
$L_\alpha(Z_N)$, which determines the fluctuations, decays as $1/Z_N^{1/\alpha}$,
exactly as the tail of $f(x)$. This explains the highly fluctuating $\tauexp$
obtained in figure~2 as being intrinsically due to the type of involved
statistics and not to some technical experimental problem. Thus, fluctuations
do not vanish as usual with the increase of the size $N$ of the statistical
sample. The sums $X_N$ retain an intrinsically large irreproducibility. This
is in contradiction with a traditional motivation for applying statistical
methods: to go beyond the irreproducibility of individual events in order 
to obtain
quasi-perfect reproducibility for large ensembles of events. However, the
generalized CLT still allows for some predictibility in the statistical sense,
since it predicts the stable form of the probability density of $Z_N$.

Physics has incorporated two new types of randomness during this century:
quantum uncertainty and deterministic chaos. It seems to us that the
non-averaging out of fluctuations in L\'evy flights can also be
recognized as an important type of randomness\footnote{J.P.~Bouchaud speaks of a 'science of irreproducible results'. B.~Mandelbrot uses the term 'wild
randomness'.}.

\end{itemize}

\section{Subrecoil laser cooling of atomic gases}

In the Arrhenius cascade (section~2), a sum of $N$ independent terms was
directly measured and the generalized CLT could be applied directly to analyze
the results. In this section, we proceed a step further by studying a richer
physical problem, called subrecoil laser cooling. In this case, L\'evy sums
---and their properties dictated by the generalized CLT--- play an essential
role, although they are not measured directly.

We introduce subrecoil laser cooling in section~3.1. In section~3.2, we show
that power law densities of time variables appear, which implies the
non-ergodicity of the process. In section~3.3, the generalized CLT is used to
get some insight on the cooling efficiency. In section~3.4, quantitative
predictions are derived, using the 'sprinkling density'. In
section~3.5, we indicate how the insight provided by the generalized CLT
enables to optimize the cooling strategy.

The starting point of the approach presented here has been 
presented in \cite{BBE94} and \cite{Bar95}. A detailed description of the
theory will appear in \cite{BBA99}.

\subsection{Subrecoil laser cooling}

Laser cooling of atomic gases is based on the momentum exchanges between
photons and atoms.  In standard ({\it not} subrecoil) laser cooling,
laser configurations and atomic transitions are carefully chosen so that 
these momentum exchanges lead to a friction force. This friction force
damps the thermal
atomic momenta $p$, thereby reducing the momentum spread (standard deviation) 
$\delta p$ of the
atomic gas, which is equivalent to reducing the effective temperature $T$ 
defined by
\be
\kB T = \delta p^2/M
\label{def_T}
\ee
where $\kB$ is the Boltzmann constant and $M$ is the mass of the atoms.
Temperatures
commonly achieved in the last ten years are in the range of a few
microkelvins, 8~orders of magnitude below room temperature. This has opened
exciting new possibilities for atomic and quantum physics and has been a key
ingredient in the realization, in 1995, 
of a new state of matter called Bose-Einstein
condensate. The 1997 Nobel prize of physics was attributed to S.~Chu,
C.~Cohen-Tannoudji and W.~Phillips for their contributions to laser cooling.

Standard laser cooling mechanisms are fundamentally limited to temperatures
larger than the so-called 'recoil temperature'. Indeed, among the momentum
exchanges between atoms and photons, some ---the ones due to spontaneous
emission--- occur in a random direction. Each spontaneous emission of a photon
by an atom thus results in an uncontrollable random recoil of the atomic 
momentum $\vec{p}$ 
by a quantity $\hbar\vec{k}$, where $\hbar\vec{k}$ is the momentum of a single
photon. Therefore, the standard deviation $\delta p$ of the atoms is expected to
be always larger than $\hbar k$. This implies (cf. Eq.~(\ref{def_T}))
laser cooling temperatures $T$ larger than the recoil temperature defined by 
$\TR = (\hbar k)^2/(\kB M)$. The recoil temperature $\TR$ is on the order of one
microkelvin for the configurations frequently used.

To sum up, the randomness of spontaneous emission, which is essential for 
the cooling since it provides a dissipative contribution
to the atomic evolution, is also harmful to the cooling 
since it implies a limit temperature. As spontaneous emission of photons by 
atoms placed in laser light seems unavoidable, the recoil temperature 
was for some time considered as an absolute limit for laser cooling.

\medskip
Subrecoil laser cooling, \ie $T<\TR$, is however possible. 
Indeed, although spontaneous emission of
photons is an intrinsically random quantum process, it can be partly
controlled. The key idea is to create a spontaneous emission rate $R(p)$ which
depends on the atomic momentum $p$ (figure~3) and which vanishes at $p=0$. 
This was first proposed and realised in 1988~\cite{AAK88,AAK89}
using a nice quantum effect called a
'dark resonance', because a resonance occurs at $p=0$ which prevents the
spontaneous emission. Today, record low
temperatures reached experimentally with dark resonances 
approach $\TR/1000$, which corresponds to a few nanokelvins only. 

\bfig{!h}
\vspace*{0.5cm}
\includegraphics[height=7cm,width=8cm,angle=0]{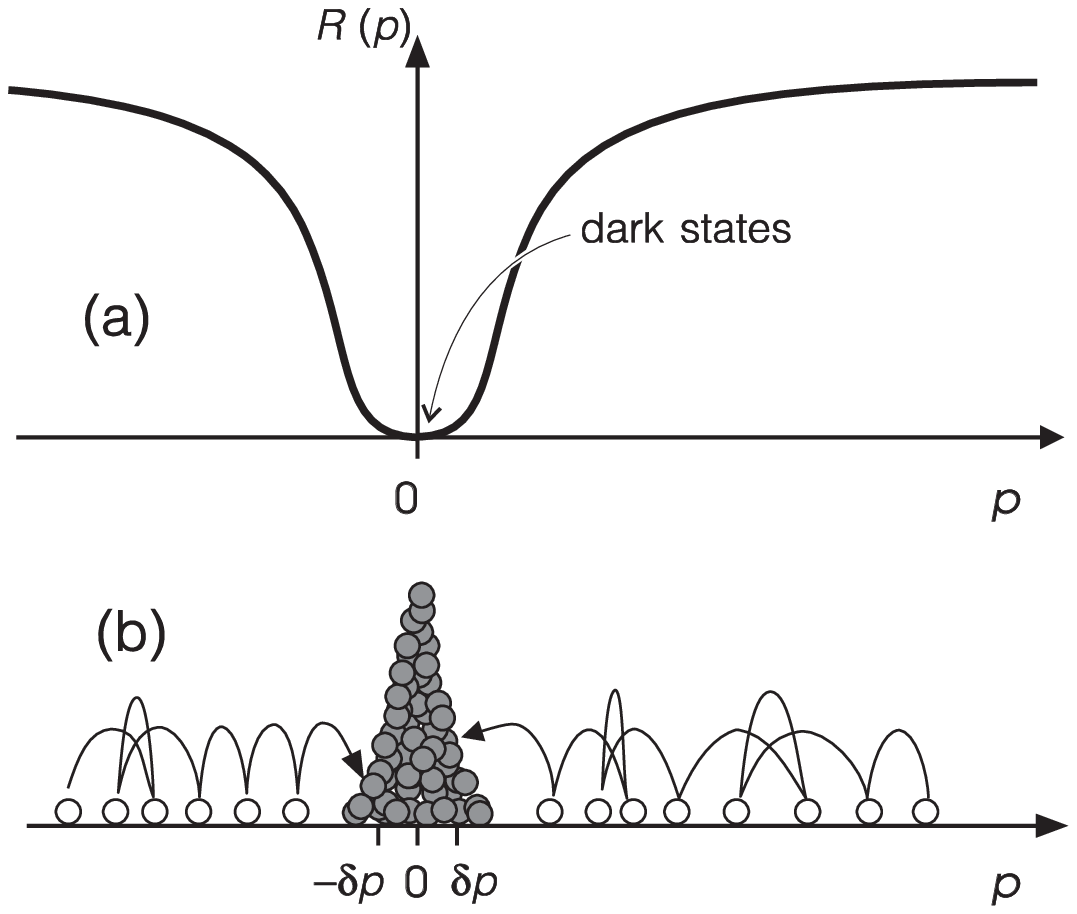}
\caption{Principle of subrecoil cooling. a) The spontaneous emission rate
$R(p)$ vanishes at momentum $p=0$. b) The atoms perform a momentum random walk
and accumulate in the vicinity of $p=0$.}
\label{fig3}
\end{figure}

One can follow the evolution of an atom with an initially non-zero momentum
$p$. The spontaneous emission rate\footnote{The spontaneous emission
rate $R(p)$ can be simply seen as a diffusion coefficient that has the peculiarity of varying with the momentum $p$.} $R(p)$ being large, the atom will
spontaneously emit photons\footnote{Before each spontaneous emission, the atom
absorbs a laser photon. The recoil effects of photon absorption,
which is a deterministic process, are not essential here and are therefore
ignored.} and therefore its
momentum will change in a random way. This random walk will eventually lead by
chance the atom in the vicinity of $p=0$ where the spontaneous emission rate
is very small (see figure~3). There, the atom stops exchanging 
momentum with photons and it
remains so-to-speak 'trapped' in what is called a 'dark state'. The
time $\tau(p)$ of residence at momentum $p$ is
\be
\tau(p) = \frac{1}{R(p)}\ ,
\label{tau_p}
\ee
the time interval between two spontaneous emissions. If this residence time is
long enough, the atom keeps the same small momentum till the end of the
experiment. If not, it emits a spontaneous photon, which starts a new momentum
diffusion process and gives to the atom  another chance to reach 
the vicinity of $p=0$.
Thus atoms accumulate in the vicinity of $p=0$ in long-lived states: a cooling
effect occurs. This cooling relies on a random walk of the atomic momentum,
unlike standard laser cooling which rests on friction forces.

\bigskip
The most important question for a cooling process is to determine the typical
momentum $\delta p$ at the end of the random walk or, equivalently the
temperature $T$ (cf. Eq.~(\ref{def_T})). It is difficult to answer it
with the usual analytical or numerical methods of atomic physics\footnote{However, in one particular case, analytical solutions based on the usual methods have been found~\cite{AlK96,SSY97}. Numerical approaches have also been developped~\cite{CBA91} using a new type of quantum simulations.} 
because very different momentum and time scales
are present in the problem. A conjecture was proposed in 1988 in which the
interaction time $\theta$, defined as the time the atoms interact with the
lasers, plays a crucial role. Consider an atom reaching a momentum $p$ such that
the residence time $\tau(p)$ is larger than the interaction time $\theta$. 
This atom
will thus keep its momentum $p$ till the end of the experiment and will be
detected with this momentum $p$. The conjecture consists in assuming that {\it
only} the atoms such that 
\be
\tau(p)\geq\theta
\label{conj1}
\ee
keep their momentum till the end of the experiment. Obviously, this can not be
strictly true: some atoms will reach a small momentum $p$ after a significant
time $t$ has elapsed from the beginning of the interaction with the lasers so
that, for them, the condition to stay at momentum $p$ would rather be $\tau(p)
\geq \theta -t$. But let us assume that condition~(\ref{conj1}) is the relevant
criterion for the trapping of atoms.  This predicts that a momentum peak will
form with a width $\delta p_\theta$ given by
\be
\tau(\delta p_\theta) \simeq \theta \ \cdot
\label{conj2}
\ee
Moreover, it can be shown that the residence time $\tau(p)$ varies as
\be
\tau(p) \propto 1/p^2 \ \cdot
\label{tau_p2}
\ee
Introducing this relation into Eq.~(\ref{conj2}) gives the conjectured
momentum scale $\delta p_\theta$ which is reached after an interaction time
$\theta$:
\be
\delta p_\theta \propto \frac{1}{\sqrt{\theta}}
\ee
or, for the temperature $T_\theta\propto (\delta p_\theta)^2$ (cf.
Eq.~(\ref{def_T})), 
\be
T_\theta \propto \frac{1}{\theta}\; \cdot
\label{conj_T}
\ee

This result is both interesting and surprising. Interestingly, it predicts
that the temperature can be reduced towards lower and lower values when
the interaction time
$\theta$ is increased. The recoil temperature limit, which arises 
in standard laser cooling from
spontaneous emission, is no more a limit here. Here
indeed, spontaneous emission is present to create a random walk that 
brings the atoms
to $p\simeq 0$, but spontaneous emission stops when the atoms 
reach a small enough momentum.
Surprisingly, there does not seem to be {\it any} limit for the cooling.

This has motivated a series of experiments with longer and longer interaction
times $\theta$. The recoil limit was first overcome in 1988, reaching
$T_\theta\simeq \TR/2$~\cite{AAK88}. Longer interaction times allowed to reach
$T_\theta\simeq \TR/40$ in 1994~\cite{BSL94,Bar95} 
and $T_\theta\simeq \TR/800$ in 1997~\cite{SHK97},
establishing a new temperature record each time. Moreover, these experiments
agree well with the conjectured temperature dependence of Eq.~(\ref{conj_T}).

However, one obviously needs a better understanding of what 
determines the temperature
$T_\theta$. A related question is the proportion of cooled atoms: can a random
walk with no driving force lead to an accumulation of all the atoms in the
vicinity of $p=0$? Is there rather only a small proportion of cooled atoms?
How does this fraction vary with the interaction time $\theta$ 
and with the number of dimensions of
the random walk? As we will see below, random walk techniques and the
generalized CLT provide answers to these questions.

\subsection{Trapping time densities and non-ergodicity}

Recently developped quantum simulations~\cite{DCM92,DZR92,Car93,CBA91} 
allow to follow the momentum random walk of
a single atom in the process of subrecoil cooling~\cite{CBA91,BBE94,Bar95}. 
An example of such a
random walk is represented in figure~4. We see how the random evolution of the
atomic momentum sometimes leads to $p\simeq 0$ states where the atom remains
for a long time because the spontaneous emission rate $R(p)=1/\tau(p)$ 
vanishes in $p=0$: this is the principle of subrecoil cooling at work.

\bfig{!h}
\vspace*{0.5cm}
\includegraphics[height=8cm,width=7cm,angle=-90]{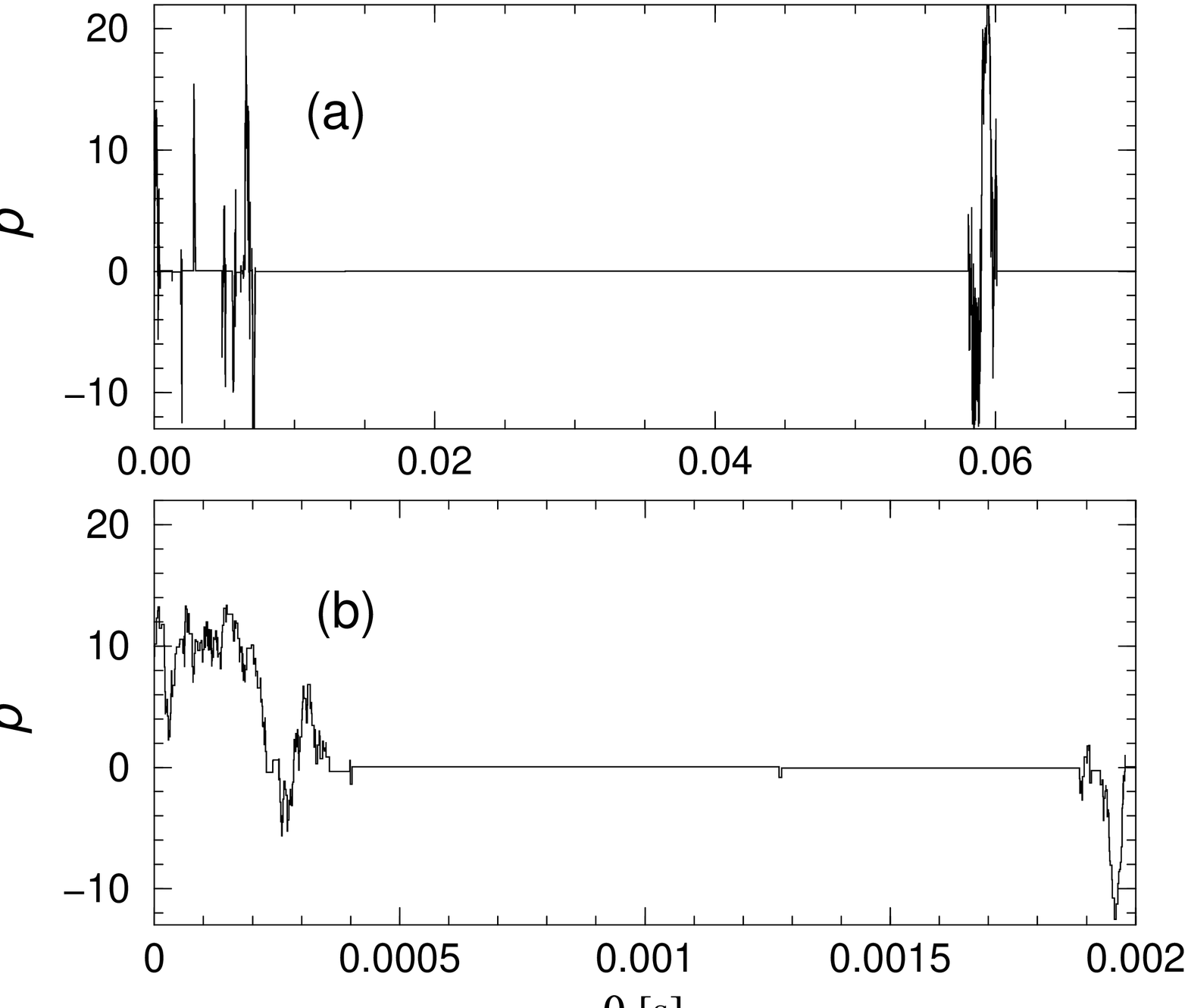}
\caption{(a) Example of a momentum random walk resulting
from a Monte-Carlo simulation of subrecoil cooling of metastable 
helium atoms.
The unit of atomic momentum $p$ is the momentum $\hbar k$ of the photons.
The zoom (b) of the beginning of the time evolution is statistically 
analogous to the evolution at large scale, a fractal property typical of a
L\'evy flight.}
\label{fig4}
\end{figure}

More importantly, figure~4 presents an interesting statistical feature that
triggered the L\'evy flight approach of laser cooling. 
The {\it single} largest
residence time $\tau_{\rm max}$ amounts to 70~\% of the total time $\theta$
while the atom has occupied 4000 different momentum states during this total 
time. Thus a single event dominates a sum of a large number of
events, which is an indication of a L\'evy flight.

Is there really a L\'evy flight in the problem? Let us estimate the
probability density $f(\tau)$ of trapping times $\tau$, 
defined as the times spent by
an atom in an interval $[-\ptrap;+\ptrap]$ called the ''trap'' of size $\ptrap$
smaller than the size $\hbar k$ of a random momentum step\footnote{Under these
conditions, the trapping times $\tau(p)$ are simply the residence times
$\tau(p)$ of Eq.~(\ref{tau_p}) and Eq.~(\ref{tau_p2}) in the region
$[-\ptrap;+\ptrap]$.}. The probability density $\rho(p)$ for an atom
reaching the trap to fall in a state of momentum $p$ can then be considered as
independent of $p$ (in one dimension): 
\be
\rho(p) \simeq \frac{1}{2\ptrap}\; \cdot
\label{rho}
\ee
The trapping time density $f(\tau)$, given by $f(\tau)\diff
\tau = \rho(p)\diff p$, is easily obtained using Eq.~(\ref{tau_p2}):
\be
f(\tau) \propto \frac{1}{\tau^{1+\alpha}} \quad {\rm with} \quad \alpha=
\frac{1}{2}\; \cdot
\label{puiss2}
\ee

Thus, if we consider as a first step that the interaction time $\theta$ is the
sum of trapping times $\tau_i$ with the density (\ref{puiss2})\footnote{We
neglect here the times $\hat{\tau}_i$ spent outside the trap. These will be
taken into account in the following section.}, the time $\theta$ is indeed a
L\'evy sum, whose behaviour is dictated by the generalized CLT with $\alpha<1$.
We have a L\'evy flight in time, which immediately accounts for the domination 
of a single trapping event in figure~4 (see section~2.3).

\medskip
There is a deep physical consequence of this L\'evy flight, {\it the absence of
ergodicity}. The ergodic hypothesis, an important ingredient in 
statistical physics, is the assumption that time averaging of a physical
quantity yields the same result as ensemble averaging. Time
averaging requires following a particle over a time much larger than all
characteristic times of the problem. This is impossible here. Indeed, 
as the time $\theta$ gets larger, larger trapping time scales (up to $\theta$)
appear and the time averaging procedure does not converge. 
This is reflected in the fact that we have a L\'evy flight on a time variable 
$\tau$\footnote{Thus, the same non-ergodic properties occur for the Arrhenius
cascade at law temperatures (see section~2).}, with infinite average 
trapping times. Thus, subrecoil cooling is a
non-ergodic process. The non-ergodicity is associated to the absence of
cooling limits. The cooling goes on for ever because larger and larger
trapping times $\tau(p)$, corresponding to lower and lower momenta $p$, 
can be reached with increasing $\theta$.

\subsection{Trapping, recycling and the generalized CLT}

The history of an atom over a time $\theta_N$ can be seen a series of $N$
trapping times $\tau_i$ interrupted by $N$ times $\hat{\tau}_i$ spent out of
the trap\footnote{Note that the initial problem
is a momentum random walk, which we treat efficiently by considering the associated random walk in time, a standard method for these problems.}. 
The times $\hat{\tau}_i$ are the usual 'first return times'. We
also call them 'recycling times' because the atoms coming out of the trap are
given another opportunity to reach the trap, they are 'recycled'. 

Thus, the interaction time $\theta_N$ writes as
\be
\theta_N = \tau_1 + \hat{\tau}_1+... + \tau_N + \hat{\tau}_N = T_N+\hat{T}_N,
\ee
where 
\be
T_N = \sum_{i=1}^{N} \tau_i
\ee
is the total trapping time,
and 
\be
\hat{T}_N = \sum_{i=1}^{N} \hat{\tau}_i
\ee
is the total recycling time. Both $T_N$ and $\hat{T}_N$ are sums of
independent variables. The application of the generalized CLT to these sums
gives in a simple way a qualitative answer for the proportion of cooled
atoms, as we discuss now. 

\medskip
Consider first the case in which the spontaneous emission rate $R(p)$ tends to 
a non-vanishing constant at large $p$. Then, at large $p$, we have a standard
random walk with a constant diffusion rate. For a 1D problem, the probability 
density $\hat{f}(\hat{\tau})$ of first return times 
$\hat{\tau}$ is known to decay at large $\hat{\tau}$ as
\be
\hat{f}(\hat{\tau}) \propto \frac{1}{\hat{\tau}^{1+\hat{\alpha}}} \quad {\rm
with}  \quad \hat{\alpha}=\frac{1}{2}\; \cdot
\ee
It thus decays exactly in the same way as $f(\tau)$. According to the 
generalized CLT, for large $N$'s, the sums $T_N$ and $\hat{T}_N$ behave as
\begin{eqnarray}
T_N & \sim & N^{1/\alpha} = N^2\ , \label{T}\\
\hat{T}_N & \sim & N^{1/\hat{\alpha}} = N^2 \, \cdot
\end{eqnarray}
Therefore, for long times (cf. large $N$'s), one has $T_N \sim \hat{T}_N$: 
the atoms spend a
finite fraction of their time in the trap and a finite fraction outside the
trap. We thus expect the proportion of cooled atoms to tend to a constant,
strictly between 0 and 1.
More elaborate calculations confirm this non-trivial result.

Consider now the case in which a friction mechanism is added to prevent the
atoms to diffuse to too large momenta $p$. This friction confines the
momentum diffusion in a finite zone. In this case, $\hat{f}(\hat{\tau})$ 
is a narrow probability density with a finite average value. 
Thus, according to the usual CLT:
\be
\hat{T}_N \simeq N \langle \hat{\tau} \rangle \, \cdot
\ee
Comparing this to Eq.~(\ref{T}) which is still valid here, one has 
\be
T_N \gg \hat{T}_N
\ee
for large $N$. This implies that all the atoms will be cooled, which is again
confirmed by more elaborate calculations.

More complicated cases can be considered by including diffusion in 2 or 3
dimensions or by including the 'Doppler effect' which modifies the rate
$R(p)$ at large $p$. In each case, the generalized CLT provides the asymptotic
behaviours of the sums $T_N$ and $\hat{T}_N$ which yield immediately the
qualitative asymptotic proportion of cooled atoms.

\subsection{Momentum distribution}

Up to now, we have presented mostly qualitative results. We want to
sketch here how some quantitative results are obtained. 

The main features of the cooling process are given by the momentum 
distribution (probability density)
${\mathcal P}(p,\theta)$ of trapped atoms at time $\theta$. This 
momentum distribution ${\mathcal P}(p,\theta)$ writes as an
integral over the times $\tl$ at which the atoms enter the trap for the last
time:
\be
{\mathcal P}(p,\theta) = \rho(p) \int_0^\theta \diff \tl \SR(\tl)
\psi(\theta-\tl|p) \; \cdot
\ee
The quantity $\rho(p)$ is the probability density for an atom entering 
the trap to
reach the momentum $p$ (in one dimension, we have seen in Eq.~(\ref{rho})
that $\rho(p)=1/(2\ptrap)$). The quantity $\SR(t)$ called here the 'sprinkling
density' is the probability density for an atom to return into the trap
at time $t$, regardless of the number of times the atom 
has entered the trap before.
The quantity $\psi(\tau|p)=\int_{\tau}^\infty f(\tau') \diff \tau'$ is
the probability that an atom remains in the trap during a time longer than
$\tau$ (where $f(\tau)$ is the trapping time density defined in
section~3.2).

The momentum distribution can be calculated explicitly. For instance, in a
simple 1D model with infinite $\pt$ and finite $\langle \hat{\tau} \rangle$,
one obtains
\be
{\mathcal P}(p,\theta) = h(\theta) {\mathcal G}(Ap\sqrt{\theta})
\label{eq.mom}
\ee
where $h(\theta)\propto \sqrt{\theta}$ is the height of the cooled peak at
$p=0$, $A$ is a constant. The function ${\mathcal G}(q)$ is given
by ${\mathcal G}(q) =1$ for $q\leq1$ and by ${\mathcal
G}(q)=1-(1-q^{-2})^{1/\alpha}$ for $q\geq1$. The width $\delta p_\theta$ of
${\mathcal P}(p,\theta)$ decays as $1/\sqrt{\theta}$, which proves the
1988 conjecture of a temperature decrease without any fundamental
limit (cf. Eq.~(\ref{conj_T})). This calculation can also be done for more complicated cases in any
dimension where it is very useful. For instance, one can study the
influence of the exponent $\beta$ in the spontaneous emission rate
$R(p)\propto p^\beta$, as described in the next section.

\medskip
The key point to obtain the momentum distribution is the calculation of the
sprinkling density $\SR(t)$. The sprinkling density is obtained 
relatively easily with a Laplace 
transform\footnote{In fact, the generalized CLT 
is not explicitly used in the derivation of Eq.~(\ref{eq.mom}).}. The result 
is interesting. If $\pt$ and
$\langle\hat{\tau}\rangle$ are finite, then $\SR(t)$ tends to a constant at
large times. This is an expected 'ergodic' result: the rate of return events
is asymptotically constant\footnote{In a Poisson process, this rate is
constant at any time scale.}. On the contrary, if $\pt$ or
$\langle\hat{\tau}\rangle$ is infinite, then $\SR(t)$ decays to 0 at large
times. This is a signature of non-ergodicity: at large times, the density of
return events go to 0 because the longer and longer $\tau_i$'s or
$\hat{\tau}_i$'s which tend to appear slow down the diffusion. Such a process
has a 'history': the measurement of $\SR(t)$ at any time tells when the
diffusion has started.

\subsection{Optimizing laser cooling with the generalized CLT}

A remarkable outcome of the usual CLT is that the statistical behaviour of
L\'evy sums $X_N$ at large $N$ is determined {\it only} by two parameters, 
$\pmo$ and $\dm$. Thus, the detailed features of $f(x)$ can be forgotten if one
is interested only in the large $N$ properties of the L\'evy sums. Similarly,
with the generalized CLT in the cases $\alpha\leq2$, only the asymptotic power
law behaviour of $f(x)$ is relevant to determine the behaviour of L\'evy sums
at large $N$. For a positive variable $x$, for instance, this power law 
is described by two parameters only, the exponent $\alpha$ and the 
prefactor of the power law. 
	This can provide a useful insight when confronted to a complex physical 
phenomenon with many parameters: 
the generalized CLT shows that the many physical parameters combine into only
two relevant statistical parameters.

Such an insight has been applied in practice to improve a subrecoil 
laser cooling mechanism called Raman cooling~\cite{KaC92}. Raman cooling, 
like the dark resonance cooling described in
section~3.1, rests on a $p$-dependent spontaneous emission rate $R(p)$
analogous to the one in figure~3a. The main difference is that the rate $R(p)$
results from the superposition of pulses that can be chosen nearly
arbitrarily. This gives flexibility to this mechanism and makes it a good case
study for cooling optimization. On the other hand, the large number of
parameters ($\simeq30$ for the initially used sequence of pulses) to be
optimized makes it necessary to find simplifying guidelines.

By carefully  using the generalized CLT, we have proposed a new very simple
sequence of pulses~\cite{RBB95}: it relies on 4~pulses only (compared to 14 initially); the
shape of the pulses is the simplest possible while the initially used pulses
were sophisticated Blackman pulses. 

The results are eloquent. With the initially used sequence of 14~pulses, the
temperature $T_\theta$ varied as $T_\theta\propto 1/\sqrt{\theta}$ 
with the interaction time
$\theta$. With the new sequence of 4~pulses, the new shape (which changes the
exponent $\beta$ of the rate $R(p)\simeq p^\beta$ from $\beta=4$ to $\beta=2$) 
leads to $T_\theta \propto
1/\theta$, a much faster cooling. Moreover, with this new shape, if the pulses
parameters (width and position) are adapted to the considered interaction time
$\theta$, one obtains an even faster cooling $T_\theta \propto 1/\theta^{4/3}$~\cite{RBB95,Rei96}.

These predictions have been successfully tested experimentally and led to
record low temperatures ($2.8\pm 0.5$~nK) for a cesium gas.
This shows how the generalized CLT can have significant practical consequences.

\section{Imperfect L\'evy flights}

We have presented in sections 2 and 3 two examples where the generalized CLT
applied perfectly. However, there are many physical cases where the
generalized CLT is useful although the conditions to apply it are not, 
strictly speaking, mathematically fulfilled. This may occur either 
because the asymptotic decay of $f(x)$ is not purely a power law or 
because $f(x)$ is a truncated power law.

Let us first discuss the 
{\it truncation problem}\footnote{In section~3, the sums
$T_N=\sum_{i=1}^{N} \tau_i$ were limited by the available interaction time
$\theta$ which is also a kind of truncation. However, this {\it truncation of
the sum itself} by an experimental parameter (here the interaction time, in
other cases the system size) does not prevent the appearance of all the
important effects of the generalized CLT; on the contrary, the fact that the 
truncation value, however large it may be, has an effect on the measured value
is a signature of the generalized CLT. The truncations dealt with in section~4
bear on the density $f(x)$ itself and imply a departure from the generalized
CLT.}, \ie the cases in which $f(x)$
decays as $1/x^{1+\alpha}$ for $x<x_0$ and decays more rapidly for $x>x_0$ so
that $\pmo$ and $\dm$ are finite. In the mathematical sense, the usual CLT
applies. However, due to the power law tail, {\it the convergence to the 
asymptotic gaussian for the
probability density of the L\'evy sums can be extremely slow}, 
being reached for $N$ typically of $10^3$ or larger~\cite{MaS94}, while 
in most cases for which the usual CLT applies, the approximate
convergence to a gaussian is obtained very rapidly, with typically 
$N\simeq 4-5$. As, in practice, one often deals
with sums of a moderately large number $N$ of terms, the behaviour of the
L\'evy sums is often dictated by the L\'evy laws for relevant $N$ values, 
while the gaussian behaviour is recovered only for irrelevantly large $N$
values. 

Second, there are {\it broad probability densities which decay only 
approximately as power
laws}. An example is provided by broad lognormal distributions, which have of
course a finite second moment. They can be rewritten as power laws
$1/x^{1+\alpha(x)}$ with a logarithmically varying exponent $\alpha(x)$. If the
logarithmic part of $\alpha(x)$ is small enough, then the generalized CLT gives
at least some qualitative guidelines for the behaviour of the L\'evy sums. We
have used such guidelines to study the tunneling of electrons through a thin
layer of insulator, a problem which has both basic and applied interests. The
striking finding related to the generalized CLT has been that the typical
current density varies by more than 200 {\it depending on the scale} 
at which it is measured~\cite{Bar97,DBB98,DHB98} (see also \cite{LaB93}),
while the typical current density should be scale independent if there were no
tails in the probability density of the current.

\subsection*{Acknowledgements}
The Institut de Physique et de Chimie des Mat\'eriaux de Strasbourg is Unit\'e
Mixte de Recherches 7504 of Centre National de la Recherche Scientifique and
of Universit\'e Louis Pasteur. The section~3 is mostly based on a long paper
(to be published) in collaboration with J.P.~Bouchaud (Saclay), 
A.~Aspect (Orsay) and C.~Cohen-Tannoudji (Paris). 
I also thank all the researchers with whom I collaborated on the quoted papers.

\end{document}